\documentstyle[12pt,epsf]{article}
\newcommand{\ds }{\displaystyle}
\newcommand{\ra}{\rightarrow}
\newcommand{\be}{\begin{equation}}
\newcommand{\ee}{\end{equation}}
\newcommand{\bea}{\begin{eqnarray}}
\newcommand{\eea}{\end{eqnarray}}
\newcommand{\ci}{\cite}
\newcommand{\bi}{\bibitem}
\newcommand{\nono}{\nonumber \\}

\newcommand{\p}{{\vec{\bf p}}}
\newcommand{\e}{{\rm e}}

\newcommand{\dd}{\partial}

\newcommand{{\bfna}}{\mbox{\boldmath$\vec{\nabla}$}}
\newcommand{\epp}{\mbox{\boldmath$\vec\epsilon$}}
\newcommand{\half}{\frac{1}{2}}

\newcommand{\jj}{\vec{\bf{j}}}
\newcommand{\FF}{\vec{\bf{F}}}
\newcommand{\vv}{\vec{\bf{v}}}
\newcommand{\JJ}{\vec{\bf{J}}}
\newcommand{\rr}{\vec{\bf{r}}}

\def\dal{\,\lower0.3ex\vbox{\hrule\hbox{\vrule\kern2pt\vbox{\kern4pt\kern4pt}
\kern2pt\vrule}\hrule}\,}

\def\L{{\cal L}}

\begin{document}

\title{\sl Ehrenfest theorem, Galilean invariance
 and nonlinear Schr\"odinger equations}
\vspace{1 true cm}
\author{G. K\"albermann$^*$
\\Soil and Water dept., Faculty of
Agriculture, Rehovot 76100, Israel}
\maketitle

\vspace{3 true cm}
\begin{abstract}

Galilean invariant Schr\"odinger equations possessing nonlinear
terms coupling the amplitude and the phase of the wave function
 can violate the Ehrenfest theorem. An example of this kind is provided.

The example leads to the proof of the theorem: A Galilean invariant 
Schr\"odinger equation derived from a lagrangian density obeys
the Ehrenfest theorem. The theorem holds for any linear or
nonlinear lagrangian.

\end{abstract}


$^*${\sl e-mail address: hope@vms.huji.ac.il}

\newpage

\section{\sl Introduction}

The linear Schr\"odinger equation for a pointlike object interacting with 
an external force derived from a potential $\ds U(\rr)$,

\bea\label{sch}
i\frac{\dd\psi}{\dd t}~=~-\frac{1}{2~m}{\bfna}^2\psi~+~U(\rr)\psi
\eea

obeys the Ehrenfest theorem\ci{ehren}.
The theorem is expressed as an equation for the velocity 
and an effective second law of Newton. The probability function for
the averaging procedure is taken to be the absolute
value squared of the wave function that solves the pertaining 
Schr\"odinger equation.

The Ehrenfest equations read

\bea\label{ehren0}
<\vv>&=&\frac{d}{dt}<\rr>\nono
<\rr>&=&\int d^3x~|\psi(\rr,t)|^2~\rr\nono
\eea

and 

\bea\label{ehren00}
m\frac{d}{dt}<\vv>&=&\FF(t)\nono
\FF(t)&=&-\int d^3x~|\psi(\rr,t)|^2~{\bfna}~U(\rr)
\eea

The theorem states that the quantum expectation values of position and
velocity of a suitable quantum system obey the classical equations of
motion.
This observation supports the choice of Born\ci{born} of the amplitude
squared as a probability weight factor. 
According to the present interpretation of quantum mechanics
\ci{ballentine}, 
the outcome of measurements of the motion of identically prepared 
pointlike systems, their location 
and speed when subjected to so-called 'external' forces, will be, on average,
the same as the sharp values of their classical counterparts.

In classical deterministic mechanics, the sources of randomness
are twofold, statistical fluctuations and extreme 
sensitivity to initial conditions.
Statistical fluctuations can be minimized by bringing the system 
in contact with a
reservoir at very low temperature, near absolute zero. The sensitivity
to initial conditions causing chaotic behavior are not omnipresent. They 
arise for nonlinear systems. Hence, 
the outcome of the measurement of a point particle is in principle, 
predetermined.
Eqs.(1) tell us that besides the statistical fluctuations, quantum
systems possess an extra source of indeterminacy, that is regulated in a very
definite way by the complex wave function. This is in some sense better for
the experimenter than classical thermal fluctuations that are hidden completely.
Even in the limit of zero temperature, 
at which the classical fluctuations vanish and
a deterministic outcome is expected, the macroscopic variables position 
and velocity are still fuzzy. The fuzziness is {\sl prescribed}
 by the Schr\"odinger equation. Hence, only averages obey
deterministic equations. 

Ehrenfest theorem can be extended to many point particle systems 
without difficulty. 
The theorem embodies Bohr's correspondence principle. As such, the dynamics
dictated by eq.(1) ought to be independent of Planck's quantum of action 
$\hbar$.
A failure of the ehrenfest theorem would carry on to the behavior of
quantum particles in external fields such as  gravitational fields. 
A system that is assumed to obey a nonlinear
Schr\"odinger equation of the kind described below, 
displays an inherent nonclassical behavior, even on the average.

In section 2 we single out the kind of nonlinearities that violate
 with Ehrenfest theorem.
In section 3 we prove a theorem that connects Galilean invariance, and 
the existence of a lagrangian density whose Euler-Lagrange equation 
is the Schr\"odinger equation, to the fulfillment of the Ehrenfest
theorem. The proof hinges upon the existence
of a lagrangian density. The existence of this density guarantees
the existence of a conserved energy and a conserved linear momentum
for the free particle, even when the particle is allowed to
self-interact.
Galilean invariance then inforces a connection between
the Ehrenfest average velocity and the conserved momentum.

\newpage
\section{\sl Ehrenfest theorem breaking}

The linear Schr\"odinger equation for a pointlike quantum particle whose
configuration wave function is $\ds~\psi(\rr)$, in
interaction with an external potential $U(\rr)$, can be derived from
the lagrangian

\bea\label{lagrange}
{\L}&=&~\frac{-i}{2}\bigg(\frac{\dd\psi^*}{\dd t}~\psi-
\frac{\dd\psi}{\dd t}~\psi^*\bigg)-\frac{1}{2~m}{\bfna}\psi^*\cdot{\bfna}\psi-
|\psi|^2~U(\rr)
\eea

The lagrangian is a real scalar. A global phase transformation on the wave 
function leaves the lagrangian invariant. The Noether current
for the symmetry becomes\ci{hagen}

\bea\label{current}
j_0&=&\psi^*~\psi\nono
\jj&=&\frac{i}{2~m}\bigg({\bfna}\psi^*~\psi-{\bfna}\psi~\psi^*\bigg)
\eea

Born\ci{born} interpreted this current as a probability current.
$\ds j_0$, the probability density, and, $\ds \jj$, the probability
flux.
For waves whose amplitude decreases fast enough at infinity, current
conservation 

\bea\label{conserv}
\frac{\dd j_0}{\dd t}~+~{\bfna}\cdot\jj=0
\eea

implies that the norm $\ds{\sl N}=\int d^3x~|\psi|^2$ is independent of time.
By rescaling the wave function to unit norm, the modified norm is in
Born's view the total probability to have the particle anywhere in space.

Ehrenfest\ci{ehren}, before Born, found that there is a way to define
a privileged location whose dynamics follows the classical equations of
motion. After Born's interpretation, it becomes clear that the definition

\bea\label{x}
<\rr>=\int d^3x~|\psi|^2~\rr
\eea

is a sound one for the averaged position variable or packet centroid. 
Derivation with respect to time of eq.(\ref{x}), application of 
eq.(\ref{conserv}), and integration by parts for well behaved wave functions
yields eq.(\ref{ehren0}) 
with 

\bea\label{vv}
<\vv(t)>~=~\int d^3x~\jj
\eea

Further derivation of the latter and use of the Schr\"odinger equation of
eq.(\ref{sch}), yields the eq.(\ref{ehren00}).

Consider now nonlinear Schr\"odinger equations. Such equations as the Gross-
Pitaevskii equation\ci{stringari} are frequently used for the description of
Bose-Einstein condensates (BEC) of Alkali gases. The equation is 
an effective equation, derivable from a field theory by taking condensate
expectation values, and it has found quite success in the description of the
BEC setup.
Starting from a linear field theory of particles,  
it is possible to arrive at a nonlinear Schr\"odinger equation
in which the interactions are of zero range to lowest order.
In this sense the effective theory is merely an approximation to the
actual complicated dynamics ruling the behavior of the particles in
interaction. As an effective theory, it has found a fair amount of
predictive power. It is logical then, to ask the question whether this
effective theory still obeys the Ehrenfest theorem. If the answer is
in the affirmative, then, the averaging procedure that lead to the
Schr\"odinger equation for the condensate did not break the connection to
classical physics. If not, then, either this breaking is a falsification
due to the approximations, or else, it really reflects an observable feature
of these systems. In our opinion, only experiment can answer this question.

Another widespread application of nonlinear Schr\"odinger equations is that
of density functional phenomenological models.
In these models, the self-interaction of the particles, 
is nonlinear in the density.
This approach yields statical and dynamical properties of systems, 
ranging from electron transport in solids\ci{weinberger}, 
electronic excitations\ci{onida}, soft condensed matter\ci{likos}  
phase transitions in liquid crystals\ci{singh}, phonons in solids \ci{fritsch,
baroni}, colloids\ci{lowen}, liquids and nuclei \ci{ ghosh}, atoms and molecules
 \ci{nagy}, quantum dots\ci{reimann}, etc.
In these theories, the Schr\"odinger equation becomes

\bea\label{sch1}
i\frac{\dd\psi}{\dd t}~=~-\frac{1}{2~m}{\bfna}^2\psi~+[O(|\psi|^2)+~U(\rr)]\psi
\eea

where $\ds O(|\psi|^2)$ is a nonlinear and sometimes nonlocal\ci{dalfovo}
 functional of the density $\rho=\ds |\psi|^2$. 
Calculating the time derivative of the velocity of eq.(\ref{vv}), with
the nonlinear term of eq.(\ref{sch1}) included, we find

\bea\label{nlehren}
m~\frac{d}{dt}<\vv(t)>~=~-\int d^3x~\rho(\rr)\half~{\bfna}~[U(\rr)+O(\rho)]
\eea

Using the trivial property

\bea\label{vanish}
\int d^3x~\rho~{\bfna}[O(\rho)]=0
\eea

we arrive at the same Ehrenfest equation as that of eq.(\ref{ehren0},
\ref{ehren00}).
Nonlinear Schr\"odinger equations with nonlinearities such as that of
 eq.(\ref{sch1}) obey the Ehrenfest relations.
Perhaps this is expected as the nonlinear terms are essentially 
classical, depending on the density $\ds \rho$ and not on the wave function 
itself.
If the {\sl external} interaction is itself
nonlinear such as $\ds U(\rho,x)$, then the Ehrenfest therem is still 
satisfied, but this time with a nonlinear force \footnote{This is analogous
to the results found in soliton models.\ci{reinisch}}

\be\label{nlf}
\FF(t)=-\int d^3x~|\psi(\rr)|^2~{\bfna}~U(\rho,\rr)
\ee

In the following we will show that nonlinear Schr\"odinger equations of a 
different type, for which the nonlinearity depends on the phase also,
do not obey the Ehrenfest equations. 
The trivial property of eq.(\ref{vanish}) is the clue to the construction of 
the desired equations. The vanishing of eq.(\ref{vanish}) is due to the 
dependence of $\ds O(\rho)$ solely on $\rho$. It is easy to show,
that an operator that depends on gradients of the phase for instance,
will yield extra terms to the Ehrenfest equations.

Consider the Schr\"odinger equation 
proposed by Doebner and Goldin\ci{doebner}, without
the diffusive term 
\footnote{A sizeable body of
literature on related topics may be traced by citations to the 
ref.\ci{doebner}}

\bea\label{doebner}
i\frac{\dd\psi}{\dd t}&=&-\frac{{\bfna}^2\psi}{2~m}+\lambda{\bfna}\cdot\bigg(
\frac{\jj}{|\psi|^2}\bigg)\psi+U(\rr)\psi\nono
\eea

where $\ds \lambda$ is a coupling constant and $\ds \jj$ is defined in 
eq.(\ref{current}). Current conservation still holds as in eq.(\ref{conserv}).
The definition of the velocity is still given by eq.(\ref{ehren0}).
Equation (\ref{doebner} is Galilean invariant. 
Galilean invariance is effected by means of the transformations
\ci{niederer,hagen,brown}

\bea\label{galilean}
\rr&\ra&\rr-\delta\vv~t\nono
\frac{\dd}{\dd t}&\ra&\frac{\dd}{\dd t}-\delta\vv\cdot{\bfna}\nono
\psi&\ra&\e^{i \phi}\psi\nono
\phi&=&\half m~(\delta\vv)^2~-~m\delta\vv\cdot\rr
\eea

with $\ds\delta\vv$ a constant velocity parameter.
Under these substitutions, the free Schr\"odinger equation is invariant
, while 
\bea\label{curr}
\frac{\jj}{|\psi|^2}\ra\frac{\jj}{|\psi|^2}+\delta\vv
\eea
Eq.(\ref{doebner}) is then Galilean invariant. 
However the Ehrenfest theorem is not satisfied

\bea\label{ehren1}
m~\frac{d}{dt}<\vv(t)>=\FF(t)+\lambda~\int~d^3x~\bfna|\psi|^2~\cdot
{\bfna}\bigg(\frac{\jj}{|\psi|^2}\bigg)
\eea

The identity (\ref{vanish}) is of no use now. The last term
in eq.(\ref{ehren1}) depends on derivatives of the amplitude and 
derivatives of the phase of the wave function, 
and not on $\ds \rho$ solely as in eq.(\ref{vanish}).

The failure to satisfy the Ehrenfest theorem may by traced back to the absence
of a lagrangian density for which eq.(\ref{doebner}) 
is its Euler-Lagrange equation of motion.
This remark will be put in the form of a theorem in the next section.

\newpage

\section{Ehrenfest theorem as a consequence of Galilean invariance}

The proof of the theorem proceeds by the following steps.
The Schr\"odinger lagrangian for a free
particle including self-interactions of any nonlinear nature,
but no explicit dependence on the space or time coordinates is introduced.
The corresponding action is then invariant under spatial 
coordinate translations.
By Noether's theorem there arises a conserved current and the physical law of
conservation of linear momentum.
The lagrangian is also demanded to be a real scalar. 
It depends on the phase of the wave function only through its derivatives.
Phase transformations will then induce the law of conservation
of probability identified as the modulus squared of the wave function.

Galilean invariance of the lagrangian then determines 
a connection between the probability current and the linear momentum. 
Finally, this connection insures the validity of the Ehrenfest theorem.

Consider a lagrangian density 

\bea\label{lag}
{\L}(\psi,\psi_t,\psi_i)&=&{\L}(S_t,S_i,R,R_t,R_i)\nono
\psi&=&R~e^{i~S}
\eea

where suffixes denote partial derivatives, with respect to time {\sl t}
 and with respect to the coordinate $ x_i$.
As we are in the nonrelativistic domain, we do not differentiate between 
covariant and contravariant indices.

Space translations are generated by the transformations
$\ds \rr\ra~\rr+\epp$, with $\ds \epp$ a constant parameter.
As the lagrangian is independent of the
coordinates, it is invariant under this transformation.
When the parameter $\ds\epp$
is promoted to be space-time dependent \ci{hagen}, the law of conservation
of linear momentum appears as a condition on the invariance of
the action for arbitrary $\epp$.

The variation of the action reads

\bea\label{action}
\delta {\sl A}&=&\delta\int~d^3x~dt~{\L}\nono
&=&\int~d^3x~dt\bigg[{\L}~{\bfna}\cdot\epp-\frac{\dd{\L}}{\dd S_t}
\epp_t\cdot{\bfna} S-\frac{\dd{\L}}{\dd R_t}\epp_t\cdot{\bfna} R\nono
&-&\frac{\dd{\L}}{\dd S_i}{\bfna} S\cdot\epp_i
-\frac{\dd{\L}}{\dd~R_i}{\bfna} R\cdot\epp_i\bigg]
\eea

The invariance of the action then engenders the law of conservation of linear
momentum

\bea\label{linear}
0&=&\frac{\dd p_j}{\dd t}+\frac{\dd T_{ij}}{\dd x_i}\nono
\eea

where

\bea\label{linearp}
\p&=&-\frac{\dd{\L}}{\dd~S_t}~{\bfna} S-\frac{\dd{\L}}{\dd R_t}
{\bfna} R\nono
&=&-\frac{\dd{\L}}{\dd~\psi_t}~{\bfna}\psi-\frac{\dd{\L}}{\dd\psi^*_t}
{\bfna}\psi^*_t\nono
T_{ij}&=&\delta_{ij}{\L}-\frac{\dd{\L}}{\dd S_i}~S_j-
\frac{\dd{\L}}{\dd R_i}~R_j\nono
&=&\delta_{ij}{\L}-\frac{\dd{\L}}{\dd\psi_i}~\psi_j-
\frac{\dd{\L}}{\dd\psi^*_i}~\psi^*_j\nono
\eea

As the lagrangian (\ref{lag}) is independent of {\sl S}, a transformation
of the form $\ds S\ra~S+\theta(x,t)$, will generate the law of conservation
of probability. Straightforward manipulations lead to

\bea\label{flux}
0&=&\frac{\dd~J_0}{\dd t}+{\bfna}\cdot\JJ\nono
\eea

where

\bea\label{fluxp}
J_0&=&-\frac{\dd{\L}}{\dd S_t}\nono
&=&i\frac{\dd{\L}}{\dd\psi^*_t}\psi^*-i\frac{\dd{\L}}{\dd\psi_t}\psi\nono
\JJ&=&-\frac{\dd{\L}}{\dd{\bfna} S}\nono
&=&i\frac{\dd{\L}}{\dd{\bfna}\psi^*}\psi^*-i\frac{\dd{\L}}{\dd{\bfna}\psi}\psi
\eea

In both equations (\ref{linearp},\ref{fluxp}) we do not specify the dynamics,
by replacing the lagrangian with the free Schr\"odinger lagrangian. The only
conditions imposed are, the very existence of a lagrangian that
is a real scalar dependent on a complex wave function and independent
of spatial coordinates except through the wave function and its derivatives.
We also assumed that there do not appear derivatives of higher order, although
 this is not an essential ingredient.
Both eqs.(\ref{linear},\ref{flux}) may be obtained also by means of the 
Euler-Lagrange equations\ci{bjorken, hill}. It is easy to check that they are
mere identities when the Euler-Lagrange equations are used.

The connection to Ehrenfest theorem now proceeds through the demand of
Galilean invariance of the Schr\"odinger equation or covariance
of the action for finite Galilean boosts. Covariance, and not invariance,
as a boost modifies the kinetic energy, and consequently changes the action.
Applying an infinitesimal galilean transformation to the lagrangian {\L}
 as specified in eqs.(\ref{galilean}), i.e. dropping the term quadratic
in the velocity, the variation of the lagrangian becomes
\footnote{ This method is straightforward for our purposes, use of the Noether
method yields a tautology.}

\bea\label{dell}
\delta{\L}&=&-\frac{\dd{\L}}{\dd S_t}~\delta\vv\cdot{\bfna}S
-\frac{\dd{\L}}{\dd R_t}~\delta\vv\cdot{\bfna}R+
\frac{\dd{\L}}{\dd{\bfna} S}~\cdot m\delta\vv\nono
&=&\delta\vv\cdot{\p}-m\delta\vv\cdot\JJ
\eea

where we have used the definitions of $\ds \p$ and $\ds\JJ$ in 
eqs.(\ref{linearp},\ref{fluxp}).
For the action to be covariant, 
which is equivalent to being invariant for infinitesimal
Galilean boosts, eq.(\ref{dell}) implies

\bea\label{demand}
{\p}=m~{\JJ}+{{\bfna}}f(\rr,t)
\eea

Galilean invariance requires the probability flux to differ from the
conserved linear momentum by at most a total divergence.
For all practical purposes both vectors are one and the same up to the
factor of the mass.

The existence of a real scalar lagrangian, even including nonlinear
self-interaction terms, and the requirement of independence on coordinates 
for the free self-interacting
is necessary and sufficient for the equivalence between 
$\ds m~\JJ$ and $\ds \p$. Necessary, as we can proceed backwards
 from eq.(\ref{demand}) and reconstruct Galilean invariance.

We are now ready to show the validity of Ehrenfest theorem for a generic
lagrangian as in eq.(\ref{lag}).

We add a scalar potential term $\ds {\L}_u=
~-U(\rr)~|\psi|^2$ to the lagrangian. Use of different 
types of potentials, such as velocity dependent vector potentials, 
do not change the conclusions. Moreover, nonlinear
potentials will generate forces as in eq.(\ref{nlf}). 

The lagrangian is no longer invariant under translations, but the action 
still is.

Equation (\ref{linear}) is now modified to

\bea\label{force}
0&=&\frac{\dd p_j}{\dd t}+\frac{\dd T_{ij}}{\dd x_i}
+\frac{\dd~U(\rr)}{\dd x_j}|\psi|^2
\eea

Integrating over space for asymptotically vanishing wave functions, we find

\bea\label{force1}
\frac{d}{dt}\int~d^3x~p_j(\rr,t)=-\int~d^3x\frac{\dd~U(\rr)}{\dd x_j}|\psi|^2
\eea

This is the second law of Newton for the 'field' momentum $\ds \p$.
It bears no relation to the Ehrenfest theorem at this stage.

Consider now the definition of $\ds <\rr>$ of eq.(\ref{ehren0}).
Using eq.(\ref{flux}) we find

\bea\label{ehren2}
\frac{d }{dt}<\rr>&=&\int~d^3x~\rr~\frac{dJ_0}{dt}\nono
&=&-\int~d^3x~\rr~{\bfna}\cdot\JJ\nono
&=&\int~d^3x~\JJ
\eea

Using eqs.(\ref{demand},\ref{force1}) we find the second law of Newton for
the centroid of the packet $\ds <\rr>$, the Ehrenfest theorem

\bea\label{newton}
m~\frac{d^2}{dt^2}<\rr>&=&m~\frac{d}{dt}~\int~d^3x~\JJ\nono
&=&\frac{d}{d t}\int~d^3x~\p\nono
&=&-\int~d^3x{\bfna}[U(\rr)]~|\psi|^2\nono
&=&<\FF>
\eea

\section{Conclusions}

We have shown that the demand of Galilean invariance on a scalar real
local lagrangian pertaining to a Schr\"odinger field, linear or nonlinear in
the wave functions, implies the validity of the Ehrenfest theorem. 
The same does not hold
for Schr\"odinger equations that are not derivable from an action principle.
Among the terms that fulfill the requirements of Galilean invariance, and
can serve as viable nonlinear self inter-actions, we can
mention the function $\ds g(q)$, with $\ds q~=R_t~+~\jj\cdot\bfna ln(R)$.

\end{document}